\documentclass[10pt,a4paper]{article}

\usepackage[utf8]{inputenc}
\usepackage[T1]{fontenc}
\usepackage{lmodern}
\usepackage{microtype}
\usepackage[top=20mm,bottom=20mm,left=20mm,right=20mm]{geometry}
\usepackage{graphicx}
\usepackage{caption}
\usepackage{subcaption}
\usepackage{amsmath,amssymb,amsfonts}
\usepackage{siunitx}
\usepackage{booktabs}
\usepackage{multirow}
\usepackage{xcolor}
\usepackage{lineno}
\usepackage[numbers,super,sort&compress]{natbib}
\usepackage[colorlinks=true,linkcolor=blue,citecolor=blue,urlcolor=blue]{hyperref}
\usepackage{setspace}
\usepackage[normalem]{ulem}

\begin{document}
\setstretch{1.5} 
\pagestyle{plain}

\title{Thicker Is Better: How Surfactant Viscosity Governs Uniform Surfactant Lung Delivery}

\author{
  Tristan Beuzelin$^{1,2}$,
  Matthieu Labousse$^{1}$,
  Marcel Filoche$^{2,*}$
}
\date{}
\maketitle
\thispagestyle{empty}

\noindent
$^{1}$Gulliver, ESPCI Paris, CNRS, PSL University, Paris, France.\\
$^{2}$Institut Langevin, ESPCI Paris, CNRS, PSL University, Paris, France.\\[4pt]
$^{*}$\href{mailto:marcel.filoche@espci.psl.eu}{marcel.filoche@espci.psl.eu}

\vspace{1em}\hrule\vspace{1em}

\begin{abstract}
\noindent
Surfactant Replacement Therapy (SRT) is a well-established intervention for Neonatal Respiratory Distress Syndrome, yet its translation to adults with Acute Respiratory Distress Syndrome (ARDS) has yielded inconsistent results in clinical trials. Using a three-dimensional, zero-dimensional biomechanical model of surfactant bolus transport through the pulmonary airway tree, we demonstrate that, while the tree-like branching architecture of the lung gives rise to discrete flow regimes that result in spatially heterogeneous and inefficient drug delivery, surfactant viscosity is a key determinant of delivery homogeneity: below an analytically derived critical viscosity, gravity causes complete surfactant exclusion from entire lung subregions regardless of dose volume. A clinically compatible tenfold to thirtyfold increase in viscosity above current formulations drives all bifurcation splitting factors toward the ideal even-split value of 0.5, enabling near-complete and spatially uniform coverage of terminal bronchioles in both infant and adult lung models. Principal component analysis across approximately 24 million simulations confirms that viscosity exerts the strongest positive influence on delivery homogeneity among all injection parameters, with dose volume being the primary determinant of efficiency. These findings provide a mechanistic rationale for reformulating exogenous surfactants with higher viscosity to improve SRT outcomes in adults.

\end{abstract}

\vspace{0.5em}\hrule\vspace{1.5em}

\section*{Introduction}

\hspace{\parindent} In the deepest parts of the human pulmonary airway system, alveoli are \qty{0.25}{mm}-diameter air sacs in which gas exchange between air and blood takes place. Their walls are coated with a pulmonary surfactant that lowers surface tension at the air–liquid interface, preventing notably alveolar collapse during exhalation\cite{ma_role_2012}. This surfactant is produced only from the last month of gestation; in premature newborns, surfactant deficiency causes a Neonatal Respiratory Distress Syndrome (NRDS). Surfactant Replacement Therapy (SRT) addresses this deficiency by instilling a liquid bolus of exogenous surfactant directly into the trachea. Driven by inspiratory airflow, the bolus propagates through the pulmonary tree and ultimately coats the alveolar walls. Since its introduction in the early 1980s, SRT has become the standard of care for NRDS and has contributed to a 50\% or greater reduction in neonatal mortality \cite{polin_surfactant_2014}, with few adverse effects\cite{walsh_aarc_2013}. Continuing research in this domain seeks to refine the instillation strategy and the chemical composition of the exogenous surfactant itself\cite{fortas_enhanced_2022,hentschel_surfactant_2020}.
\\\\
The success of SRT in premature infants provided a strong rationale for extending it to adults with Acute Respiratory Distress Syndrome (ARDS), a condition accounting for approximately 10\% of all intensive care unit admissions and 23\% of patients on mechanical ventilation, with a mortality rate of around 40\%\cite{bellani_epidemiology_2016}. Yet despite decades of controlled trials in both animals and humans, no study has demonstrated a clear, reproducible clinical benefit\cite{meng_effect_2019}. The reasons for these divergent outcomes are manifold and continue to be debated\cite{amigoni_surfactants_2017, lee_surfactant_2026, dushianthan_pulmonary_2023}, spanning differences in administration protocols and the biological heterogeneity of ARDS itself. Yet, one possibility has received comparatively little attention: that the failure is partly mechanical, rooted in the physics of fluid transport through an adult-sized airway tree.\\
\\
A zero-dimensional (0D) model of fluid transport inside a 3D symmetric lung architecture~\cite{filoche_three-dimensional_2015} showed that the equations of fluid transport in an adult pulmonary tree leads to inhomogeneous and inefficient delivery. Two mechanisms were identified: volume lost to the trailing film lining the conducting airways, and uneven plug splitting at airway bifurcations. Subsequent numerical studies confirmed that gravity is a primary determinant of this inhomogeneity, and that both the administered volume and the injection flow rate critically affect treatment outcomes\cite{kazemi_taskooh_transport_2019, grotberg_did_2017}. Yet, the rheology of the instilled liquid has never been considered as a control parameter. All commercially available exogenous surfactant formulations share similarly low viscosity\cite{lu_kinematic_2009, thai_rheology_2019} (Table~\ref{tab:rheology}), leaving unexplored whether adjusting this property could improve delivery.\\
\\
Here we show, through a model of bolus transport\cite{filoche_three-dimensional_2015, kazemi_taskooh_transport_2019}, across a wide clinical parameter space that viscosity is the dominant modifiable determinant of delivery homogeneity. We derive a critical viscosity, $\mu_c$, below which gravity-driven plug splitting renders entire subregions of the adult lung inaccessible. Raising viscosity above $\mu_c$ restores homogeneous delivery without a commensurate loss of efficiency. A principal component analysis of 24~million simulations confirms that viscosity is the leading variable governing delivery homogeneity across both infant and adult lungs. Notably, commercially available viscosity of exogenous surfactant lies below~$\mu_c$. These results suggest that a clinically compatible increase in surfactant viscosity could substantially improve current protocols, opening an unexplored formulation pathway for surfactant therapy.

\section*{Results}

\subsection*{Viscosity controls the spatial reach of surfactant delivery}

\begin{figure}[h!]
\centering
\includegraphics[width=\textwidth]{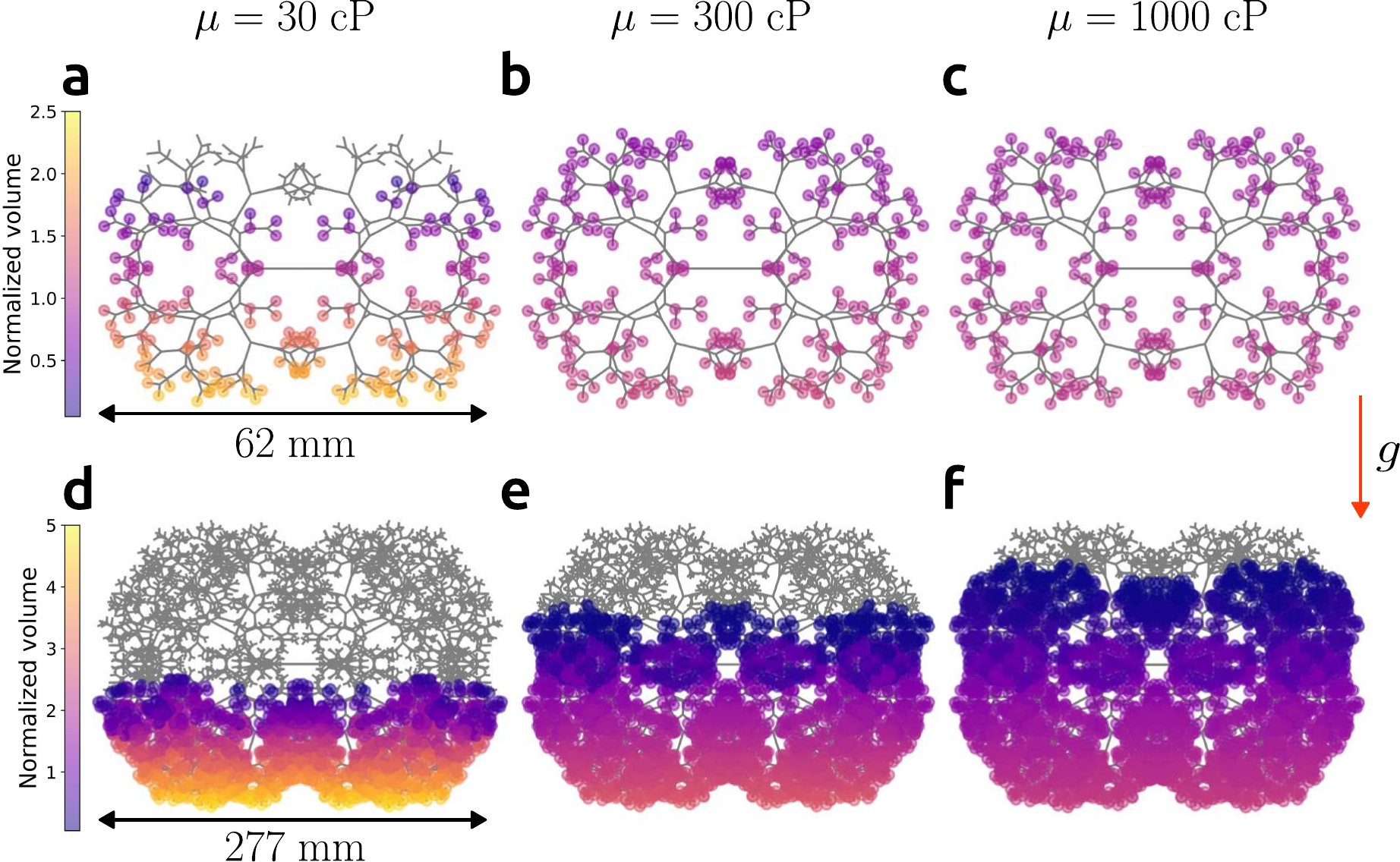}
\caption{\textbf{Surfactant delivery maps for infant and adult lungs at three viscosity values.}
Heat maps show normalised delivery (volume reaching node / initial bolus volume) at each terminal node. Top row: infant lung (\textbf{a}–\textbf{c}); bottom row: adult lung (\textbf{d}–\textbf{f}). Columns: $\mu = \qty{3e-2}{Pa\cdot s}$ (\textbf{a},\,\textbf{d}), $\qty{3e-1}{Pa\cdot s}$ (\textbf{b},\,\textbf{e}), $\qty{1}{Pa\cdot s}$ (\textbf{c},\,\textbf{f}). Trees viewed from below; the red arrow indicates gravity (supine position). Infant: $V_{\rm init} = \qty{1}{mL}$, $Q = \qty{1}{mL/s}$. Adult: $V_{\rm init} = \qty{100}{mL}$, $Q = \qty{100}{mL/s}$. At standard viscosity (\textbf{d}), the upper half of the adult lung receives no surfactant; coverage is restored progressively as viscosity increases (\textbf{e},\,\textbf{f}).}
\label{fig:trees}
\end{figure}

\begin{figure}[h]
\centering
\includegraphics[width=\textwidth]{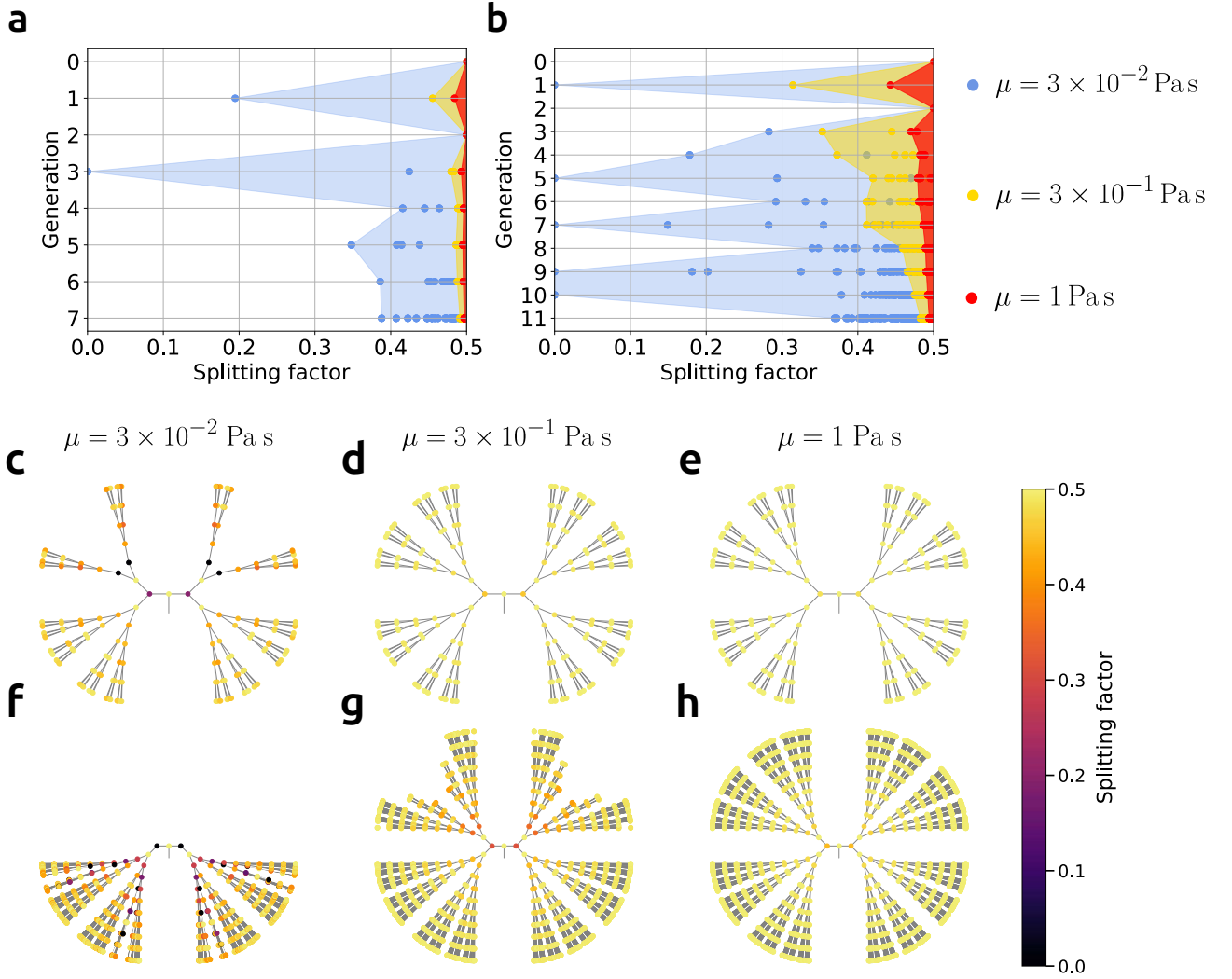}
\caption{
\textbf{Splitting factors across airway generations.}
\textbf{a},~\textbf{b}, Each point is the splitting factor~$\alpha$ at one bifurcation; generation~$0$ is the bifurcation immediately distal to the trachea. $\alpha = 1/2$ (colorbar to the right) corresponds to perfectly even splitting, whereas $\alpha = 0$ means that the plug enters the lower daughter airway in its entirety, blocking every downstream branch of the upper subtree. \textbf{a},~Infant lung. \textbf{b},~Adult lung. Colours denote the three viscosities of Fig.~\ref{fig:trees}. In the adult lung, several generations show $\alpha \approx 0$ at standard viscosity; raising the viscosity shifts all splitting factors towards~$1/2$. \textbf{c}--\textbf{h}, Flattened representation of the airway tree for the same three viscosities, showing the value and the position of every splitting factor within the binary tree. A branching path either reaches its terminal node or stops early, because the splitting factor vanishes (black dot) or because the plug is lost to lubrication (the branch ends without a black dot). The upper half of each graph corresponds to the upper half of the lung, gravity being defined for a supine patient. \textbf{c}--\textbf{e},~Infant lung. \textbf{f}--\textbf{h},~Adult lung.
}
\label{fig:alphas}
\end{figure}

\hspace{\parindent} We simulated bolus transport through symmetric Weibel-type airway trees\cite{weibel_architecture_1962, weibel_morphometry_1963} which consists of a dichotomous branching tree. Each generation $n$ of airways of radius $a_n$ ($n=0,1,\ldots $) divides at a bifurcation~$n$ in daughter branches of internal radius $a_{n+1}$ with a constant reduction in size ratio $\lambda=a_{n+1}/a_n=2^{-1/3}$. (\textcolor{black}{See Methods}). Using lubrication theory and pressure balance equations, the propagation of a liquid plug of density, $\rho$, viscosity, $\mu$, into in the pulmonary airway tree is described as a succession of two-step processes: first, propagation of the surfactant-laden plug inside an individual airway. The plug loses part of its volume by leaving behind a trailing film coating the airway. Second, splitting of the plug at the subsequent bifurcation between the two daughter airways. The input parameters of such a model are the lung architecture and the initial conditions of the injection. The latter include the volume administered to the patient, the entry flow rate, the position of the patient, and the rheology of the surfactant. We investigate two cases:  a 1-kg premature infant lung (lung tree with 8~generations, $2^{8} = 256$ terminal bronchioles) and an adult lung (lung tree with 12~generations, $2^{12} = 4096$ terminal bronchioles), for three viscosity values: $\mu = \qty{3e-2}{Pa\cdot s}$ (standard viscosity in current medical practice, {\it e.g.} Curosurf), $\qty{3e-1}{Pa\cdot s}$, and $\qty{1}{Pa\cdot s}$. Injection volume, $V_{\rm init}$, and injection flow-rate, $Q_{\rm init}$, matched current typical clinical protocols: \qty{1}{mL} at \qty{1}{mL/s} (infant) and \qty{100}{mL} at \qty{100}{mL/s} (adult), with the patient supine---lying on their back---.\\
\\
In the infant lung, surfactant reaches the majority of terminal airways even at Curosurf viscosity, though with a dorsal bias (Fig.~\ref{fig:trees}a). A tenfold viscosity increase suffices to supply all 256 terminal bronchioles (Fig.~\ref{fig:trees}b–c). The adult lung tells a different story: at Curosurf viscosity, the entire upper half of the lung receives no surfactant whatsoever (Fig.~\ref{fig:trees}d). The upper lung becomes progressively accessible only as viscosity increases (Fig.~\ref{fig:trees}e–f), with near-homogeneous coverage at $\qty{1}{Pa\cdot s}$.\\
\\
The main quantities governing the homogeneity of the terminal distribution of surfactant are the \emph{splitting factors} at the bifurcations of the pulmonary airway tree. At each bifurcation, this quantity is defined as the ratio of the plug volume propagating into the upward daughter airway (relative to gravity) to the plug volume that reached the bifurcation. The splitting factor follows from a pressure balance across the plug menisci (\textcolor{black}{see Methods}), and takes the generic form at a given bifurcation---which labels are omitted for brevity---:
\begin{equation}\label{eq:alpha_split}
\alpha = \frac{1}{2}\!\left(1 - \frac{X\sin\theta\,\sin\varphi}
{1 - X\cos\theta\,\sin\gamma}\right), \quad
X = \frac{2\lambda^4\tilde{V}}{\lambda^2\,\mathrm{Re} + 16 \tilde{V}}
\cdot \left(\frac{\rho g a^2}{\mu U}\right) \,,
\end{equation}
where $\theta$ is the bifurcation half-angle, $\varphi$ and $\gamma$ are the roll and pitch angles of the patient relative to gravity of magnitude $g$,  $\mathrm{Re} = \rho aU/\mu$ is the Reynolds number, $\rho g a^2/(\mu U)$ is a dimensionsless number---also called Jeffreys number. In these equations, $a$ and $U$ are the radius and the fluid speed before the bifurcation, respectively, and $\tilde{V} = V(\mu)/(\pi a^3)$ is the dimensionless plug volume at the bifurcation. Given that the coating is viscosity-dependent, $\tilde{V}( \mu)$ and $U(\mu)$ inherits this dependence (\textcolor{black}{see Methods}). All these quantities depend on the position of the bifurcation in the airways tree, the previous splits and coating history.\\
\\
We rationalized quantitatively the difference of fluid propagation between the infant and the adult case by considering the splitting factors across generations in the respective airway trees (Fig.~\ref{fig:alphas}). At each bifurcation, for a symmetric tree, the value of the splitting factor $\alpha$ ranges in $[0, 1/2]$, with $\alpha = 1/2$ representing perfect equal splitting. In the infant lung, factors are already close to $1/2$ at commercially used viscosity. In the adult lung, multiple generations show $\alpha \approx 0$, meaning entire upward daughter subtrees receive nothing. Increasing viscosity narrows the splitting factors distribution toward $1/2$, turning these blocked subtrees into accessible ones.

\subsection*{A critical viscosity governs upper-lung access}

\begin{figure}[h]
\centering
\includegraphics[width=\linewidth]{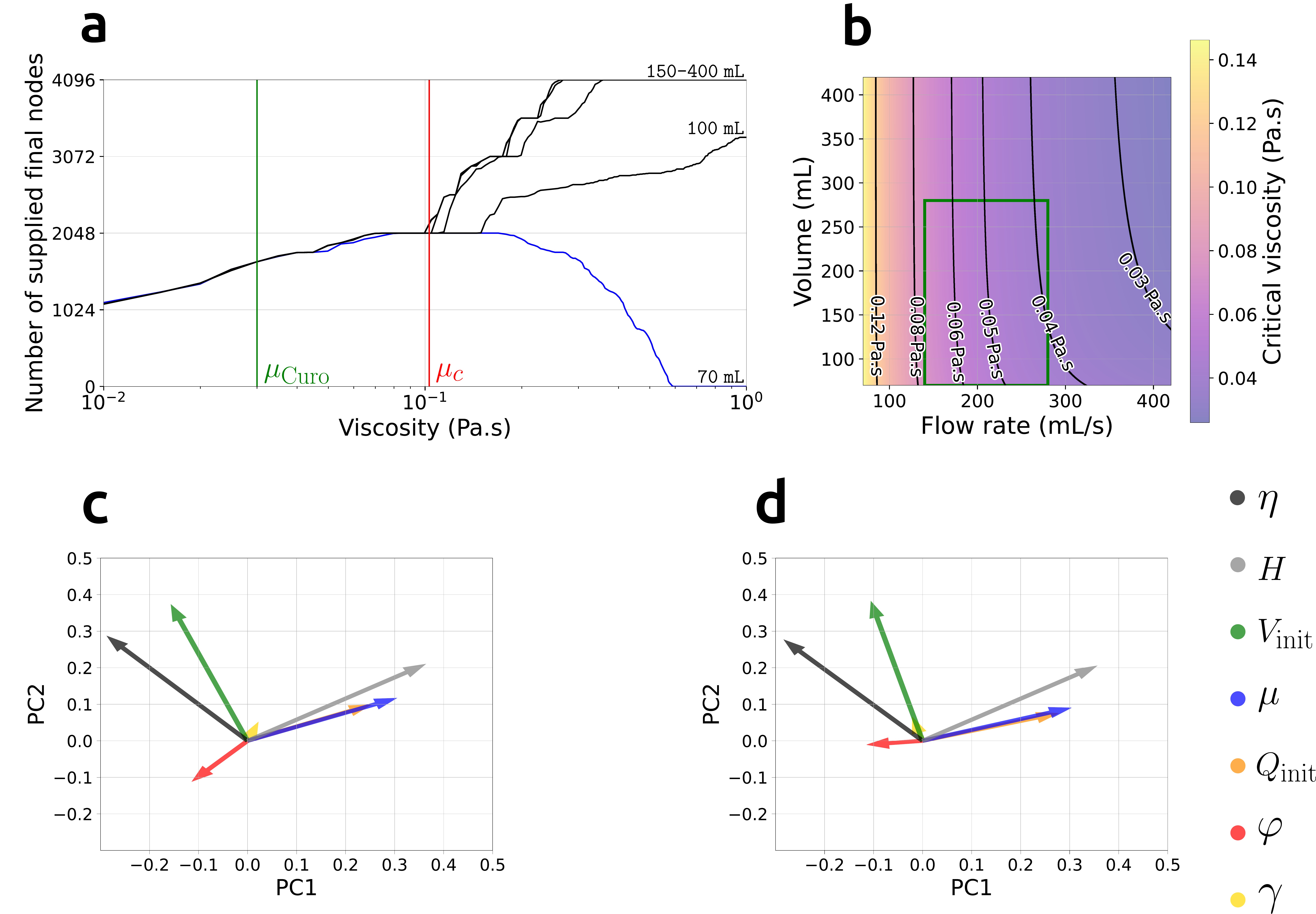}
\caption{\textbf{Critical viscosity for upper-lung access.}
\textbf{a}:~Number of supplied terminal nodes versus viscosity for several dose volumes (adult supine lung). The red vertical line marks $\mu_c = \qty{1.03e-1}{Pa\cdot s}$ (Eq.~\eqref{eq:muc}) and the green vertical line marks the Curosurf viscosity $\mu_c = \qty{3e-2}{Pa \cdot s}$. Below $\mu_c$, only 50\% of nodes are reachable regardless of dose.
\textbf{b}:~Phase diagram of $\mu_c$ in the (flow rate, dose volume) plane. The $\mu_c = \qty{3.4e-2}{Pa\cdot s}$ line is Curosurf viscosity. The green frame represents the limits of what injection parameters have been used in clinical studies \cite{kesecioglu_exogenous_2009, walsh_aarc_2013, spragg_treatment_2003}. 
\textbf{c-d}:~Principal component analysis of 24 million simulations. Contribution of each injection parameter and performance index to the two leading principal components (PC1 and PC2). Vectors are scaled by the square root of explained variance. Viscosity has the largest loading on the homogeneity axis in both infant and adult lungs.}
\label{fig:visco_crit}
\end{figure}

\hspace{\parindent} In a supine patient, the first bifurcation---connecting generations $n=0$ and $1$---lies in a horizontal plane and splits the plug evenly. The second bifurcation is the first to be asymmetrically oriented ($\varphi = \pi/2$, $\gamma = 0$); one daughter branch ascends while the other descends. At this bifurcation, the term $ 16 \tilde{V}$ in Eq~\eqref{eq:alpha_split} surpasses $\lambda^2 {\rm Re}$ by typically one order of magnitude such that Eq.~\eqref{eq:alpha_split} reduces to a simpler form:
\begin{equation}\label{eq:Xsimp}
X_1 \simeq \frac{\lambda^4}{8} \cdot \left(\frac{\rho g a_1^2}{\mu U_1}\right) \, ,    
\end{equation}
where $a_1 = \lambda a_0$ is the first-generation airway radius and $U_1(\mu)$ is the speed plug in the first generation airway. In the absence of gravity ($X=0$), splitting would be equal and an equivalent strategy is revealed by the presence of the ratio $g/\mu_c$ in Eq.~\eqref{eq:Xsimp}: restoring an homogeneous delivery can be either done by suppressing gravity at fixed viscosity or equivalently by increasing viscosity at fixed gravity---a more practical option in the context of intensive care units. Setting $\alpha = 0$ in Eq.~\eqref{eq:alpha_split} for the second bifurcation and solving for $\mu$ yields a critical viscosity, $\mu_c$, satisfying the implicit equation [\textcolor{black}{See the exact formula, Eq.~\eqref{eq:muc}}, in Methods]:
\begin{equation}\label{eq:mucsimp}
\frac{\mu_c U_1(\mu_c)}{\rho g a_1^2}\simeq \frac{\lambda^4}{8}\sin\theta\, .
\end{equation}
Below $\mu_c$, the plug bypasses the upper lung entirely; above it, upper-lung delivery resumes. Solving numerically the Eq.~\eqref{eq:muc} (Methods) gives $\mu_c = \qty{1.03e-1}{Pa\cdot s}$, approximately three times the viscosity of Curosurf. The existence of this critical viscosity is consistent with the inhomogeneity evidenced in~Fig.~\ref{fig:trees}d–e: complete upper-lung exclusion at
$\qty{3e-2}{Pa\cdot s}$, partial recovery at $\qty{3e-1}{Pa\cdot s}$. For the infant lung model, with the corresponding geometry and injection parameters (tracheal radius $a_0 = \qty{0.2}{cm}$, $2\theta = 90^{\circ}$,
$\lambda = 2^{-1/3}$, $\rho=10^3~{\rm kg/m^3}$, $V_{\rm init} = \qty{1}{mL}$, $Q_{\rm init} = \qty{1}{mL/s}$), Eq.~\eqref{eq:muc} gives $\mu_c = \qty{1.7e-2}{Pa\cdot s}$. Depending on whether the fluid viscosity is smaller or larger than $\mu_c$, the physical picture differs \\
\\
Below the critical viscosity---as is the case for Curosurf---the splitting factor vanishes at generation 1, limiting surfactant delivery to at most 50\% of terminal nodes in the supine posture, regardless of the injected dose (Fig.~\ref{fig:visco_crit}a). This sharp break point reflects the threshold-driven nature of plug splitting in fractal airway trees. Above the critical viscosity, the number of reached nodes becomes dose-dependent, as higher viscosity promotes thicker coating on airway walls. At low injection volumes (70 mL), a large fraction of the dose is retained by the coating, reducing the number of supplied nodes. However, the amount of liquid absorbed by the coating saturates as viscosity increases \textcolor{black}{[see Eq.~\eqref{eq:coating} in Methods]}, so that at sufficiently large doses, enough liquid overcomes the coating and the number of supplied nodes grows up to full lung filling.\\
\\
The critical viscosity depends on the injection parameters, mainly flow rate and to a lesser extent dose volume (Fig.~\ref{fig:visco_crit}b).  The question that arises is then the following: which volume and flow rate would be necessary to split the plug of current commercial exogenous surfactants almost evenly at the second bifurcation? Figure~\ref{fig:visco_crit}b answers to this question: reaching the same threshold at Curosurf viscosity would require flow rates and volumes outside clinical practices~\cite{walsh_aarc_2013, polin_surfactant_2014}.

\subsection*{Viscosity dominates homogeneity across the full parameter space}

\begin{figure}[h]
\centering
\includegraphics[width=\textwidth]{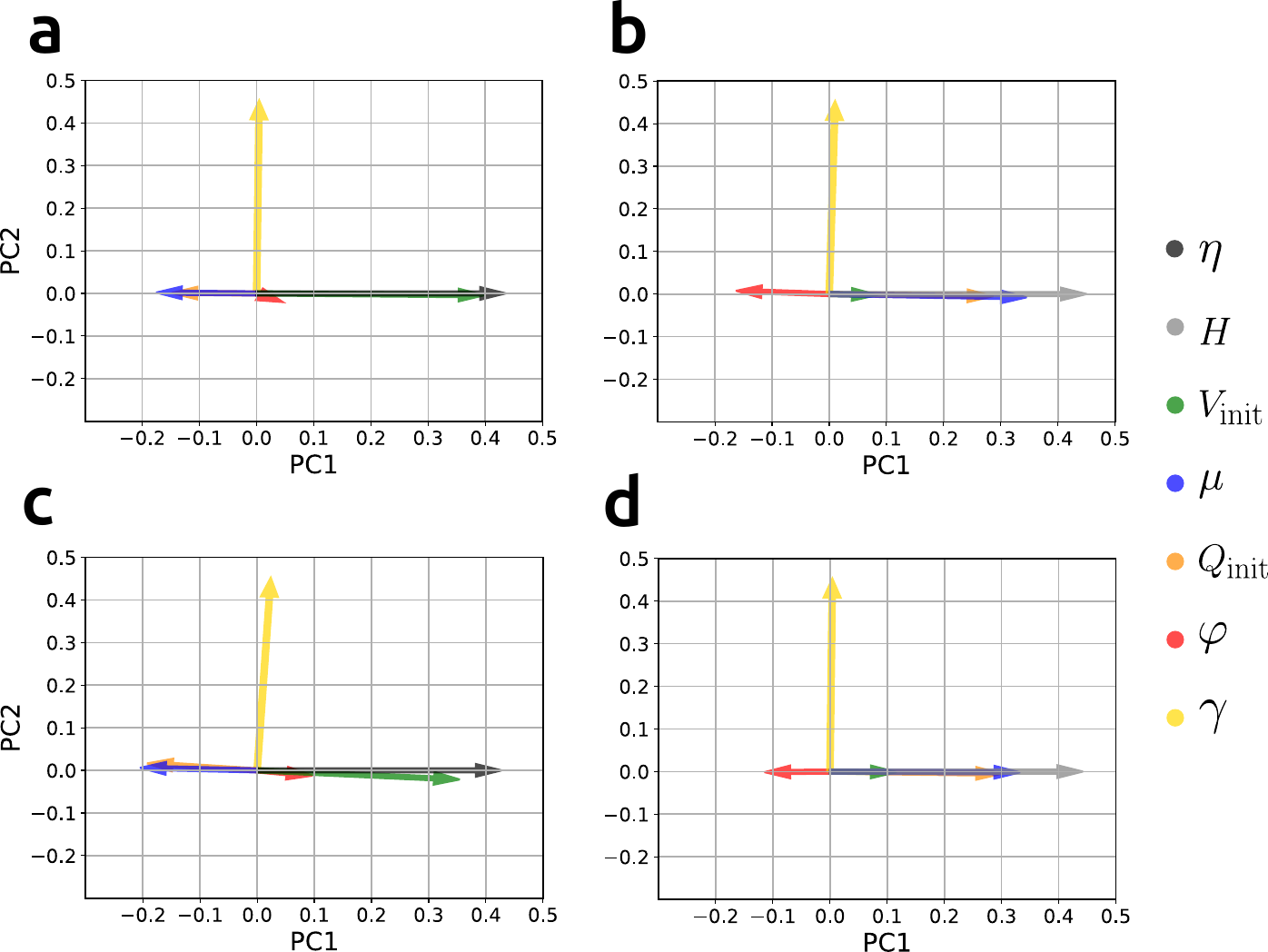}
\caption{\textbf{Principal component analysis of 24 million
simulations.}
Contribution of each injection parameter and performance index to the two leading principal components. Vectors are scaled by the square root of explained variance. Viscosity has the largest loading on the homogeneity axis in both infant and adult lungs.}
\label{fig:PCA}
\end{figure}

\hspace{\parindent} Five control parameters govern the system: initial volume, flow rate, viscosity, roll angle, and pitch angle. To assess their relative importance and mutual correlations, each parameter is varied over 30 values over several orders of magnitude, yielding $30^5 \approx 24$ million simulations for the adult and infant lung model. The computational efficiency of the model makes such a large-scale exploration tractable. We quantified outcomes with two indices: efficiency $\eta$, i.e., the fraction of bolus reaching the acinar region and homogeneity $H$ defined as the inverse of the coefficient of variation of the terminal distribution (\textcolor{black}{see Methods}). After log-transformation and
standardisation, we perform a Principal Component Analysis (PCA).\\
\\
The results are unambiguous (Extended Data, Fig.~\ref{fig:PCA}): viscosity emerges as the dominant driver of homogeneity across both infant and adult lungs, dominating the leading principal component. Dose volume enhances both efficiency and homogeneity by reducing fractional coating losses along conducting airways. Patient orientation has the smallest effect on both metrics---it redistributes surfactant spatially without changing overall acinar deposition. Notably, increasing viscosity improves homogeneity without a compensatory loss in efficiency.

\section*{Discussion}

\hspace{\parindent} The persistent failure of SRT in adults has been attributed to
disease heterogeneity, administration protocols, and biological
differences from neonatal ARDS. Our investigation demonstrates another fundamental mechanical constraint: the viscosity of current surfactant
formulations is too low to overcome gravitational plug splitting in
adult-sized airway: half the lung is inaccessible regardless of how much surfactant is instilled.\\
\\
A critical viscosity is found $\mu_c$ to emerge from the balance between two competing forces at each bifurcation: gravity, which drives the plug into the lower daughter airway, and viscous resistance, which opposes this redistribution. In neonatal airways, smaller radii weaken gravitational effects relative to viscous ones, rendering standard formulations adequate. In adult airways, gravity dominates viscous resistance with current formulation. This finding provides an imperative: viscosity must be increased threefold to restore symmetric splitting.\\
\\
Several limitations should guide future work. First, real lungs are asymmetric; the Weibel model overestimates the regularity of bifurcation geometry. Second, the coating law [Eq.~\eqref{eq:coating}] is derived for Newtonian fluids in the lubrication limit; viscoelastic or shear-thinning formulations would require an extended model. Third, a viscosity increase must not suppress the adsorption kinetics that make surfactant therapeutically active. These limitations do not affect the result of our investigation: the existence of a critical viscosity. A hard delivery threshold follows from the topology of the airway tree and the competition between gravity and viscosity, and is independent of the specific geometry or of the coating model assumed.

\section*{Methods}

\subsection*{Lung geometry}

\begin{figure}[h]
\centering
  \centering
  \includegraphics[width=\linewidth]{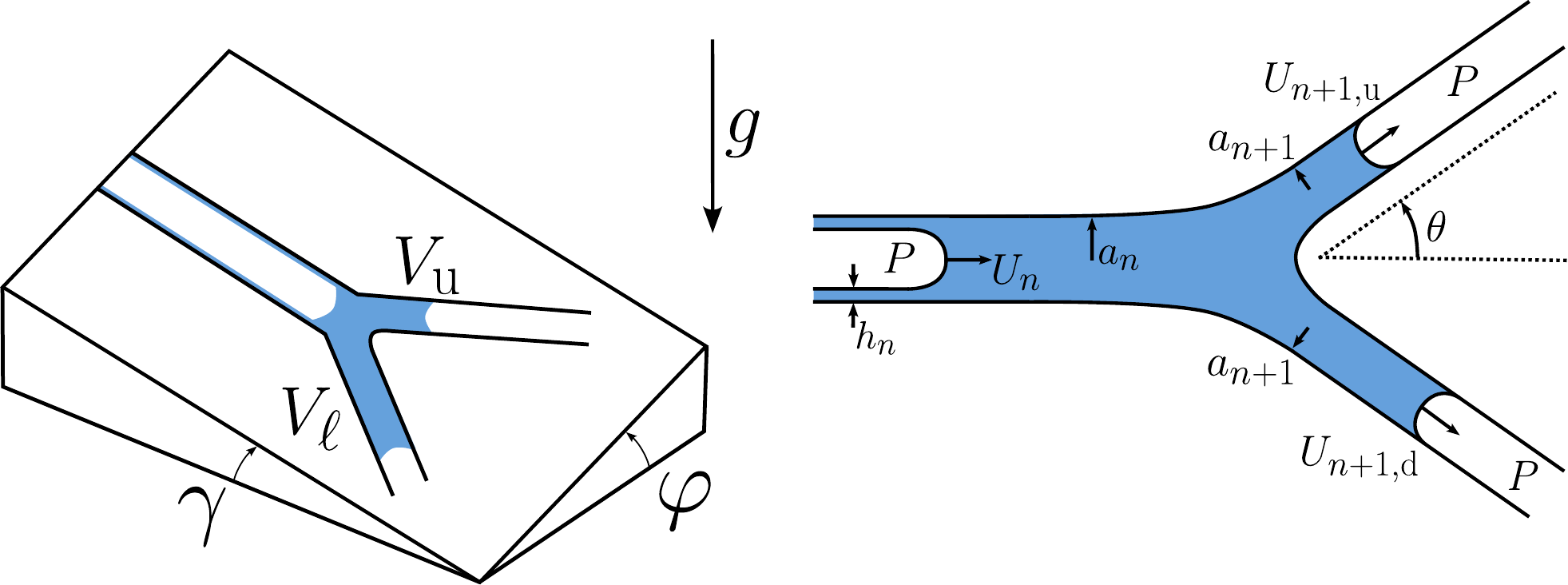}
\caption{\textbf{Bifurcation geometry and pressure balance.}
\textbf{Left}:~Roll angle $\varphi$ and pitch angle $\gamma$ define the orientation of the bifurcation plane relative to gravity.
\textbf{Right}:~Schematic of the pressure balance governing plug splitting (adapted from ref.\cite{filoche_three-dimensional_2015}).}
\label{fig:bifurcation}
\end{figure}

\hspace{\parindent} Two symmetric bifurcating trees were constructed following Weibel's morphometric model\cite{weibel_architecture_1962,weibel_morphometry_1963} (see Fig.~\ref{fig:bifurcation}), in which consecutive bifurcation planes are perpendicular. Airway radii scale as $a_{n+1}= \lambda\, a_{n}$ with $\lambda = 2^{-1/3}$ (volume-preserving branching), lengths as $L_n = 6a_n$, and daughter airways subtend an angle $2\theta = 90^{\circ}$ at each bifurcation. The infant lung model has tracheal radius $a_0 = \qty{2}{mm}$ and 8 generations ($2^8 = 256$~terminal bronchioles). The adult lung has $a_0 = \qty{8.6}{mm}$ and 12 generations ($2^{12} = 4096$ terminal bronchioles).

\subsection*{Plug propagation: coating}

\hspace{\parindent} A Newtonian liquid plug of volume $V_0$, viscosity $\mu$, density $\rho$ and surface tension $\sigma$ propagates through an airway of radius $a$, of length $L$, and at mean velocity $U$ (see Fig.~\ref{fig:bifurcation})b. The relative trailing film thickness is given by an extension of Bretherton
theory\cite{halpern_boundary_1994,Halpern_theoretical_1998}:
\begin{equation}\label{eq:coating}
\frac{h}{a} = 0.36\left(1 - e^{-2\,\mathrm{Ca}^{0.523}}\right),
\qquad \mathrm{Ca} = \frac{\mu U}{\sigma}\,.
\end{equation}
The volume entering the next bifurcation is:
\begin{equation}
V_1 = V_0 - \left[1 - \left(1-\frac{h}{a}\right)^{\!2}\right]\pi a^2 L\,.
\end{equation}

\subsection*{Plug splitting at bifurcations}

\hspace{\parindent} At each bifurcation, a pressure balance is applied across the plug menisci, incorporating the Laplace pressure drop, Poiseuille viscous drops with gravity body forces, and a momentum term for velocity change at the junction similar to the mathematical model of Filoche \textit{et al.}~\cite{filoche_three-dimensional_2015}. At each bifurcation, we define a splitting factor $\alpha$ as a dimensionless number such that the plug volume $V_1$ arriving at the bifurcation is divided between the two daughters airways into a volume $V_u=\alpha V_1$ going upward and a volume $V_d = (1-\alpha) V_1$ going downward. By construction, the volume~$V_d$ is thus always larger than $V_u$ and we have $0 \le \alpha \le 1/2$, see Eq.~\eqref{eq:alpha_split}, in the main text. The bifurcation orientation relative to gravity is parameterised by a roll angle $\varphi$ and a pitch angle $\gamma$ (Fig.~\ref{fig:bifurcation}). For a supine patient, the first asymmetric bifurcation (second generation) has $\varphi = \pi/2$ and $\gamma = 0$. Surfactant physical properties: surface tension $\sigma = \qty{3e-2}{N/m}$, density $\rho = \qty{e3}{kg/m^3}$\cite{bernhard_commercial_2000}. Default viscosity $\mu = \qty{3e-2}{Pa\cdot s}$ (Curosurf)\cite{thai_rheology_2019}. At the second bifurcation, with $V_1$ and $U_1$ the surfactant volume entering the second bifrucation and its velocity, the exact formula used to compute $\mu_c$ reads:
\begin{equation}\label{eq:muc}
\tilde{V}_1(\mu_c)\!\left(2\lambda^4 \rho g \sin\theta - 16\,\mu_c \frac{U_1(\mu_c)}{a_1^2}\right) = \lambda^2 \rho \frac{U_1^2(\mu_c)}{a_1}\, .
\end{equation}

\subsection*{Performance metrics}

\hspace{\parindent} The efficiency index $\eta$ is the percentage of the initial bolus reaching the pulmonary acini:
\begin{equation}
\eta = 100 \times \frac{\sum_{i=1}^N V_{{\rm f},i}}{V_{\rm init}}\,,
\end{equation}
where $V_{{\rm f},i}$ is the volume reaching terminal node $i$ and $N$ the total number of terminal nodes. The homogeneity index $H$ is the reciprocal of the coefficient of variation of the terminal distribution:
\begin{equation}
H = \left(\frac{N\sum_i V_{{\rm f},i}^2}{\left(\sum_i
V_{{\rm f},i}\right)^{\!2}} - 1\right)^{\!-1/2}\,.
\end{equation}
Both $H$ and $\eta$ increase monotonically with delivery quality; $H \to \infty$ for a perfectly uniform distribution.

\subsection*{Parametric study and PCA}

\hspace{\parindent} A full-factorial grid of $30^5 \approx 24$ million parameter combinations was investigated. Dose volume ranged from \qty{1}{mL} to \qty{6}{mL} (infant) and \qty{70}{mL} to \qty{420}{mL} (adult), covering the range from standard Curosurf dosing (1.25–\qty{2.5}{mL/kg}) to higher-concentration preparations such as Surfaxin (\qty{5.8}{mL/kg})\cite{filoche_three-dimensional_2015, walsh_aarc_2013}. Flow rate ranged from \qty{1}{mL\cdot s^{-1}\cdot kg^{-1}} to \qty{6}{mL\cdot s^{-1}\cdot kg^{-1}}, based on values reported in the literature\cite{noauthor_ventilation_2000, pleil_physics_2021, tsangaris_effect_2007, kazemi_taskooh_transport_2019}. Roll and pitch angles each varied from \ang{0} to \ang{90}, encoding the full range of clinically relevant patient positions. Viscosity spanned from \qty{1e-3}{Pa\cdot s} (water) to \qty{0.1}{Pa\cdot s}.

To place $H$ and $\eta$ on comparable scales, both were log-transformed: $\widetilde{H} = \ln(1+H)$, $\tilde\eta = \ln(1+\eta)$. Each metric, $\widetilde{H}$ and $\tilde\eta$ was then standardised to zero mean and unit variance before PCA. Six analyses were performed (infant/adult $\times$ both metrics on Fig.~\ref{fig:visco_crit} / $\eta$ only and $H$ only on Extended Data Fig.~\ref{fig:PCA}); the two most contributing PCA eigenvectors are shown scaled by the square root of their fraction of total explained variance. Others eigenvectors either do not contain homogeneity or efficiency contributions or explain a very minimal portion of the total variance.

\section*{Data and code availability}

Code is deposited at \url{https://doi.org/10.5281/zenodo.20628563}. All data used for this article can be regenerated with the model script.

\section*{Acknowledgements}

This work is funded by the ANR Grant INHALE (ANR-23-CE45-0009-03).

\section*{Author contributions}
T.B.: conceptualisation, methodology, formal analysis,
visualisation, writing – original draft.
M.L.: conceptualisation, supervision, writing – review \& editing.
M.F.: conceptualisation, funding acquisition, supervision, writing –
review \& editing.

\section*{Competing interests}
The authors declare no competing interests.

\bibliography{sn-bibliography}

@article{meng_effect_2019,
	title = {Effect of surfactant administration on outcomes of adult patients in acute respiratory distress syndrome: a meta-analysis of randomized controlled trials},
	volume = {19},
	issn = {1471-2466},
	shorttitle = {Effect of surfactant administration on outcomes of adult patients in acute respiratory distress syndrome},
	url = {https://bmcpulmmed.biomedcentral.com/articles/10.1186/s12890-018-0761-y},
	doi = {10.1186/s12890-018-0761-y},
	language = {en},
	number = {1},
	urldate = {2025-09-21},
	journal = {BMC Pulmonary Medicine},
	author = {Meng, Shan-Shan and Chang, Wei and Lu, Zhong-Hua and Xie, Jian-Feng and Qiu, Hai-Bo and Yang, Yi and Guo, Feng-Mei},
	month = dec,
	year = {2019},
	pages = {9},
}

@article{filoche_three-dimensional_2015,
	title = {Three-dimensional model of surfactant replacement therapy},
	volume = {112},
	issn = {0027-8424, 1091-6490},
	url = {https://pnas.org/doi/full/10.1073/pnas.1504025112},
	doi = {10.1073/pnas.1504025112},
	language = {en},
	number = {30},
	urldate = {2025-09-11},
	journal = {Proceedings of the National Academy of Sciences},
	author = {Filoche, Marcel and Tai, Cheng-Feng and Grotberg, James B.},
	month = jul,
	year = {2015},
	pages = {9287--9292},
}

@article{dushianthan_pulmonary_2023,
	title = {Pulmonary {Surfactant} in {Adult} {ARDS}: {Current} {Perspectives} and {Future} {Directions}},
	volume = {13},
	copyright = {https://creativecommons.org/licenses/by/4.0/},
	issn = {2075-4418},
	shorttitle = {Pulmonary {Surfactant} in {Adult} {ARDS}},
	url = {https://www.mdpi.com/2075-4418/13/18/2964},
	doi = {10.3390/diagnostics13182964},
	language = {en},
	number = {18},
	urldate = {2025-09-11},
	journal = {Diagnostics},
	author = {Dushianthan, Ahilanandan and Grocott, Michael P. W. and Murugan, Ganapathy Senthil and Wilkinson, Tom M. A. and Postle, Anthony D.},
	month = sep,
	year = {2023},
	pages = {2964},
}

@article{halpern_theoretical_1998,
	title = {A theoretical study of surfactant and liquid delivery into the lung},
	volume = {85},
	issn = {8750-7587, 1522-1601},
	url = {https://www.physiology.org/doi/10.1152/jappl.1998.85.1.333},
	doi = {10.1152/jappl.1998.85.1.333},
	language = {en},
	number = {1},
	urldate = {2025-09-11},
	journal = {Journal of Applied Physiology},
	author = {Halpern, D. and Jensen, O. E. and Grotberg, J. B.},
	month = jul,
	year = {1998},
	pages = {333--352},
}

@article{spragg_treatment_2003,
	title = {Treatment of {Acute} {Respiratory} {Distress} {Syndrome} with {Recombinant} {Surfactant} {Protein} {C} {Surfactant}},
	volume = {167},
	issn = {1073-449X, 1535-4970},
	url = {https://www.atsjournals.org/doi/10.1164/rccm.200207-782OC},
	doi = {10.1164/rccm.200207-782OC},
	language = {en},
	number = {11},
	urldate = {2026-01-18},
	journal = {American Journal of Respiratory and Critical Care Medicine},
	author = {Spragg, Roger G. and Lewis, James F. and Wurst, Wilhelm and Häfner, Dietrich and Baughman, Robert P. and Wewers, Mark D. and Marsh, James J.},
	month = jun,
	year = {2003},
	pages = {1562--1566},
}

@article{kesecioglu_exogenous_2009,
	title = {Exogenous {Natural} {Surfactant} for {Treatment} of {Acute} {Lung} {Injury} and the {Acute} {Respiratory} {Distress} {Syndrome}},
	volume = {180},
	issn = {1073-449X, 1535-4970},
	url = {https://www.atsjournals.org/doi/10.1164/rccm.200812-1955OC},
	doi = {10.1164/rccm.200812-1955OC},
	language = {en},
	number = {10},
	urldate = {2026-01-18},
	journal = {American Journal of Respiratory and Critical Care Medicine},
	author = {Kesecioglu, Jozef and Beale, Richard and Stewart, Thomas E. and Findlay, George P. and Rouby, Jean-Jacques and Holzapfel, Laurent and Bruins, Peter and Steenken, Edmee J. and Jeppesen, Ole K. and Lachmann, Burkhard},
	month = nov,
	year = {2009},
	pages = {989--994},
}

@article{amigoni_surfactants_2017,
	title = {Surfactants in {Acute} {Respiratory} {Distress} {Syndrome} in {Infants} and {Children}: {Past}, {Present} and {Future}},
	volume = {37},
	issn = {1173-2563, 1179-1918},
	shorttitle = {Surfactants in {Acute} {Respiratory} {Distress} {Syndrome} in {Infants} and {Children}},
	url = {http://link.springer.com/10.1007/s40261-017-0532-1},
	doi = {10.1007/s40261-017-0532-1},
	language = {en},
	number = {8},
	urldate = {2026-01-18},
	journal = {Clinical Drug Investigation},
	author = {Amigoni, Angela and Pettenazzo, Andrea and Stritoni, Valentina and Circelli, Maria},
	month = aug,
	year = {2017},
	pages = {729--736},
}

@article{ma_role_2012,
	title = {The {Role} of {Surfactant} in {Respiratory} {Distress} {Syndrome}},
	volume = {06},
	issn = {1874-3064},
	url = {https://openrespiratorymedicinejournal.com/VOLUME/06/PAGE/44/},
	doi = {10.2174/1874306401206010044},
	language = {en},
	number = {1},
	urldate = {2026-01-19},
	journal = {The Open Respiratory Medicine Journal},
	author = {Ma, Christopher Cheng-Hwa and Ma, Sze},
	month = jul,
	year = {2012},
	pages = {44--53},
}

@article{walsh_aarc_2013,
	title = {{AARC} {Clinical} {Practice} {Guideline}. {Surfactant} {Replacement} {Therapy}: 2013},
	volume = {58},
	copyright = {https://www.liebertpub.com/nv/resources-tools/text-and-data-mining-policy/121/},
	issn = {0020-1324},
	shorttitle = {{AARC} {Clinical} {Practice} {Guideline}. {Surfactant} {Replacement} {Therapy}},
	url = {https://www.liebertpub.com/doi/10.4187/respcare.02189},
	doi = {10.4187/respcare.02189},
	language = {en},
	number = {2},
	urldate = {2026-01-20},
	journal = {Respiratory Care},
	author = {Walsh, Brian K and Daigle, Brandon and DiBlasi, Robert M and Restrepo, Ruben D},
	month = feb,
	year = {2013},
	pages = {367--375},
}

@article{hentschel_surfactant_2020,
	title = {Surfactant replacement therapy: from biological basis to current clinical practice},
	volume = {88},
	issn = {0031-3998, 1530-0447},
	shorttitle = {Surfactant replacement therapy},
	url = {https://www.nature.com/articles/s41390-020-0750-8},
	doi = {10.1038/s41390-020-0750-8},
	language = {en},
	number = {2},
	urldate = {2026-01-20},
	journal = {Pediatric Research},
	author = {Hentschel, Roland and Bohlin, Kajsa and Van Kaam, Anton and Fuchs, Hans and Danhaive, Olivier},
	month = aug,
	year = {2020},
	pages = {176--183},
}

@article{polin_surfactant_2014,
	title = {Surfactant {Replacement} {Therapy} for {Preterm} and {Term} {Neonates} {With} {Respiratory} {Distress}},
	volume = {133},
	issn = {0031-4005, 1098-4275},
	url = {https://publications.aap.org/pediatrics/article/133/1/156/68334/Surfactant-Replacement-Therapy-for-Preterm-and},
	doi = {10.1542/peds.2013-3443},
	language = {en},
	number = {1},
	urldate = {2026-01-20},
	journal = {Pediatrics},
	author = {Polin, Richard A. and Carlo, Waldemar A. and {COMMITTEE ON FETUS AND NEWBORN} and Papile, Lu-Ann and Polin, Richard A. and Carlo, Waldemar and Tan, Rosemarie and Kumar, Praveen and Benitz, William and Eichenwald, Eric and Cummings, James and Baley, Jill},
	month = jan,
	year = {2014},
	pages = {156--163},
}

@article{lee_surfactant_2026,
	title = {Surfactant therapy for the treatment of acute respiratory distress syndrome: time to revisit?},
	volume = {140},
	issn = {8750-7587, 1522-1601},
	shorttitle = {Surfactant therapy for the treatment of acute respiratory distress syndrome},
	url = {https://journals.physiology.org/doi/10.1152/japplphysiol.00850.2025},
	doi = {10.1152/japplphysiol.00850.2025},
	language = {en},
	number = {1},
	urldate = {2026-01-22},
	journal = {Journal of Applied Physiology},
	author = {Lee, Kevin G. and Greendyk, Richard A. and Goligher, Ewan C.},
	month = jan,
	year = {2026},
	pages = {303--321},
}

@article{bellani_epidemiology_2016,
	title = {Epidemiology, {Patterns} of {Care}, and {Mortality} for {Patients} {With} {Acute} {Respiratory} {Distress} {Syndrome} in {Intensive} {Care} {Units} in 50 {Countries}},
	volume = {315},
	issn = {0098-7484},
	url = {http://jama.jamanetwork.com/article.aspx?doi=10.1001/jama.2016.0291},
	doi = {10.1001/jama.2016.0291},
	language = {en},
	number = {8},
	urldate = {2026-01-23},
	journal = {JAMA},
	author = {Bellani, Giacomo and Laffey, John G. and Pham, Tài and Fan, Eddy and Brochard, Laurent and Esteban, Andres and Gattinoni, Luciano and Van Haren, Frank and Larsson, Anders and McAuley, Daniel F. and Ranieri, Marco and Rubenfeld, Gordon and Thompson, B. Taylor and Wrigge, Hermann and Slutsky, Arthur S. and Pesenti, Antonio and {for the LUNG SAFE Investigators and the ESICM Trials Group}},
	month = feb,
	year = {2016},
	pages = {788},
}

@article{grotberg_did_2017,
	title = {Did {Reduced} {Alveolar} {Delivery} of {Surfactant} {Contribute} to {Negative} {Results} in {Adults} with {Acute} {Respiratory} {Distress} {Syndrome}?},
	volume = {195},
	issn = {1073-449X, 1535-4970},
	url = {https://www.atsjournals.org/doi/10.1164/rccm.201607-1401LE},
	doi = {10.1164/rccm.201607-1401LE},
	language = {en},
	number = {4},
	urldate = {2026-01-23},
	journal = {American Journal of Respiratory and Critical Care Medicine},
	author = {Grotberg, James B. and Filoche, Marcel and Willson, Douglas F. and Raghavendran, Krishnan and Notter, Robert H.},
	month = feb,
	year = {2017},
	pages = {538--540},
}

@article{fortas_enhanced_2022,
	title = {Enhanced {INSURE} ({ENSURE}): an updated and standardised reference for surfactant administration},
	volume = {181},
	issn = {0340-6199, 1432-1076},
	shorttitle = {Enhanced {INSURE} ({ENSURE})},
	url = {https://link.springer.com/10.1007/s00431-021-04301-x},
	doi = {10.1007/s00431-021-04301-x},
	language = {en},
	number = {3},
	urldate = {2026-01-27},
	journal = {European Journal of Pediatrics},
	author = {Fortas, Feriel and Loi, Barbara and Centorrino, Roberta and Regiroli, Giulia and Ben-Ammar, Rafik and Shankar-Aguilera, Shivani and Yousef, Nadya and De Luca, Daniele},
	month = mar,
	year = {2022},
	pages = {1269--1275},
}

@phdthesis{kazemi_taskooh_transport_2019,
	title = {Transport of complex fluids in the human pulmonary airway system},
	school = {Université Paris-Saclay},
	author = {Kazemi Taskooh, Alireza},
	year = {2019},
}

@book{weibel_morphometry_1963,
	address = {Berlin, Heidelberg},
	title = {Morphometry of the {Human} {Lung}},
	copyright = {http://www.springer.com/tdm},
	isbn = {978-3-642-87555-7 978-3-642-87553-3},
	url = {http://link.springer.com/10.1007/978-3-642-87553-3},
	doi = {10.1007/978-3-642-87553-3},
	language = {en},
	urldate = {2026-01-28},
	publisher = {Springer Berlin Heidelberg},
	author = {Weibel, Ewald R.},
	year = {1963},
}

@article{weibel_architecture_1962,
	title = {Architecture of the {Human} {Lung}: {Use} of quantitative methods establishes fundamental relations between size and number of lung structures},
	volume = {137},
	issn = {0036-8075, 1095-9203},
	shorttitle = {Architecture of the {Human} {Lung}},
	url = {https://www.science.org/doi/10.1126/science.137.3530.577},
	doi = {10.1126/science.137.3530.577},
	language = {en},
	number = {3530},
	urldate = {2026-01-28},
	journal = {Science},
	author = {Weibel, Ewald R. and Gomez, Domingo M.},
	month = aug,
	year = {1962},
	pages = {577--585},
}

@article{bernhard_commercial_2000,
	title = {Commercial versus {Native} {Surfactants}: {Surface} {Activity}, {Molecular} {Components}, and the {Effect} of {Calcium}},
	volume = {162},
	issn = {1073-449X, 1535-4970},
	shorttitle = {Commercial versus {Native} {Surfactants}},
	url = {https://www.atsjournals.org/doi/10.1164/ajrccm.162.4.9908104},
	doi = {10.1164/ajrccm.162.4.9908104},
	language = {en},
	number = {4},
	urldate = {2026-01-28},
	journal = {American Journal of Respiratory and Critical Care Medicine},
	author = {Bernhard, Wolfgang and Mottaghian, Jasmin and Gebert, Andreas and Rau, Gunnar A. and von der HARDT, Horst and Poets, Christian F.},
	month = oct,
	year = {2000},
	pages = {1524--1533},
}

@article{thai_rheology_2019,
	title = {On the rheology of pulmonary surfactant: {Effects} of concentration and consequences for the surfactant replacement therapy},
	volume = {178},
	issn = {09277765},
	shorttitle = {On the rheology of pulmonary surfactant},
	url = {https://linkinghub.elsevier.com/retrieve/pii/S0927776519301614},
	doi = {10.1016/j.colsurfb.2019.03.020},
	language = {en},
	urldate = {2026-01-28},
	journal = {Colloids and Surfaces B: Biointerfaces},
	author = {Thai, L.P.A. and Mousseau, F. and Oikonomou, E.K. and Berret, J.-F.},
	month = jun,
	year = {2019},
	pages = {337--345},
}

@article{pleil_physics_2021,
	title = {The physics of human breathing: flow, timing, volume, and pressure parameters for normal, on-demand, and ventilator respiration},
	volume = {15},
	issn = {1752-7155, 1752-7163},
	shorttitle = {The physics of human breathing},
	url = {https://iopscience.iop.org/article/10.1088/1752-7163/ac2589},
	doi = {10.1088/1752-7163/ac2589},
	number = {4},
	urldate = {2026-02-02},
	journal = {Journal of Breath Research},
	author = {Pleil, Joachim D and Ariel Geer Wallace, M and Davis, Michael D and Matty, Christopher M},
	month = oct,
	year = {2021},
	pages = {042002},
}

@article{noauthor_ventilation_2000,
	title = {Ventilation with {Lower} {Tidal} {Volumes} as {Compared} with {Traditional} {Tidal} {Volumes} for {Acute} {Lung} {Injury} and the {Acute} {Respiratory} {Distress} {Syndrome}},
	volume = {342},
	issn = {0028-4793, 1533-4406},
	url = {http://www.nejm.org/doi/abs/10.1056/NEJM200005043421801},
	author = {{The Acute Respiratory Distress Syndrome Network}},
	doi = {10.1056/NEJM200005043421801},
	language = {en},
	number = {18},
	urldate = {2026-02-02},
	journal = {New England Journal of Medicine},
	month = may,
	year = {2000},
	pages = {1301--1308},
}

@article{tsangaris_effect_2007,
	title = {The effect of exogenous surfactant in patients with lung contusions and acute lung injury},
	volume = {33},
	issn = {0342-4642, 1432-1238},
	url = {https://link.springer.com/10.1007/s00134-007-0597-z},
	doi = {10.1007/s00134-007-0597-z},
	language = {en},
	number = {5},
	urldate = {2026-02-02},
	journal = {Intensive Care Medicine},
	author = {Tsangaris, I. and Galiatsou, E. and Kostanti, E. and Nakos, G.},
	month = may,
	year = {2007},
	pages = {851--855},
}

@article{lu_kinematic_2009,
	title = {Kinematic viscosity of therapeutic pulmonary surfactants with added polymers},
	volume = {1788},
	copyright = {https://www.elsevier.com/tdm/userlicense/1.0/},
	issn = {00052736},
	url = {https://linkinghub.elsevier.com/retrieve/pii/S0005273609000066},
	doi = {10.1016/j.bbamem.2009.01.005},
	language = {en},
	number = {3},
	urldate = {2026-08-28},
	journal = {Biochimica et Biophysica Acta (BBA) - Biomembranes},
	author = {Lu, Karen W. and Pérez-Gil, Jesús and Taeusch, H. William},
	month = mar,
	year = {2009},
	pages = {632--637},
}

@article{halpern_boundary_1994,
	title = {Boundary {Element} {Analysis} of the {Time}-{Dependent} {Motion} of a {Semi}-infinite {Bubble} in a {Channel}},
	volume = {115},
	copyright = {https://www.elsevier.com/tdm/userlicense/1.0/},
	issn = {00219991},
	url = {https://linkinghub.elsevier.com/retrieve/pii/S0021999184712022},
	doi = {10.1006/jcph.1994.1202},
	language = {en},
	number = {2},
	urldate = {2026-08-28},
	journal = {Journal of Computational Physics},
	author = {Halpern, D. and Gaver, D.P.},
	month = dec,
	year = {1994},
	pages = {366--375},
}

\clearpage

\begin{table}[h]
\centering
\caption{\textbf{Rheological properties of commercially available exogenous surfactants.}}
\label{tab:rheology}
\begin{tabular}{lcc}
\toprule
\textbf{Surfactant} & \textbf{Density (\unit{kg\cdot m^{-3}})} &
\textbf{Viscosity (\unit{Pa\cdot s})}\\
\midrule
Curosurf  & $1000$ & $[\qty{17.2e-3}{}, \qty{25.2e-3}{}]$ \\
Survanta  & $1000$ & $[\qty{5e-3}{}, \qty{50e-3}{}]$ \\
\bottomrule
\end{tabular}
\end{table}

\end{document}